\documentclass[11pt]{article}
\usepackage{graphicx,amsmath,amsfonts,amssymb,dcolumn}
\usepackage{arev}

\begin{document}

\title{\textbf{Affine gauge theory of gravity and its reduction to the Riemann-Cartan geometry}\footnote{Work presented by R.~F.~S.~at Recent Developments in Gravity (NEB-14) June 8-11, Ioannina 2010, Greece.}}
\author{\textbf{R.~F.~Sobreiro$^1$}\thanks{sobreiro@if.uff.br} \ and \textbf{V.~J.~Vasquez Otoya$^2$}\thanks{victor.vasquez@ifsudestemg.edu.br}\\\\
\textit{{\small $^1$UFF $-$ Universidade Federal Fluminense, Instituto de F\'{\i}sica, Campus da Praia Vermelha,}} \\
\textit{{\small Avenida General Milton Tavares de Souza s/n, 24210-346, Niter\'oi, RJ, Brasil.}}\\
\textit{{\small $^2$IFSEMG $-$ Instituto Federal de Educa\c{c}\~ao, Ci\^encia e Tecnologia,}} \\
\textit{{\small Rua Bernardo Mascarenhas 1283, 36080-001, Juiz de Fora, MG, Brasil}}}
\date{}
\maketitle

\begin{abstract}
We discuss a possible framework for the construction of a quantum gravity theory where the principles of QFT and general relativity can coexist harmonically. Moreover, in order to fix the correct gauge group of the theory we study the most general one, the affine group and its natural reduction to the orthogonal group. The price we pay for simplifying the geometry is the presence of matter fields associated with the nonmetric degrees of freedom of the original setup. The result is independent of the starting theory.
\end{abstract}

\section{Introduction}\label{intro}

The description of gravity as a gauge theory based on the Lorentz group \cite{Utiyama:1956sy,Kibble:1961ba} puts gravity at the same level as the other fundamental interactions. As extra features of the gauge approach we can point out the fact that spinors can minimally couple to gravity through torsion and there is no need to linearize the fields to define a local gauge symmetry. However, the standard Einstein-Hilbert action shows itself to be not renormalizable even in this approach. To face the renormalizability problem there were proposed several alternatives by extending the Einstein-Hilbert action to more general ones \cite{Lovelock:1971yv,Mardones:1990qc,Zanelli:2005sa}. However, their quantum consistency still lacks, giving space to alternatives beyond the action level. One of them is to consider more general groups for the gauge symmetry that also allows the identification of the gauge space with the spacetime cotangent bundle, in particular we recall the so called metric-affine gravities \cite{Hehl:1976my,Hehl:1994ue}, based on the affine gauge group $A(d,\mathbb{R})$. This class of theories also requires generalized actions to account the quantum divergences and extra degrees of freedom. The present article is in fact devoted to the study of some geometrical properties of this class of theories and the development of a possible scenario for its quantization.

Gravity as geometry, where spacetime encodes the dynamical variables, is in fact incompatible with quantum field theory which considers coordinates as parameters. To avoid this ambiguity we start by improving the ideas developed in \cite{Sobreiro:2007pn,Sobreiro:2007cm,Sobreiro:2010ji} in which gauge symmetry and spacetime are considered as independent from each other. This means that the gauge space is not identified with the cotangent bundle from the beginning. In fact, all quantum fields considered cannot be identified with gravity fields unless a dynamical mass parameter emerges in the theory. This property is ensured by considering fields that are of dimension 1 and thus cannot be identified with vierbeins nor premetrics. Once the mass parameters emerge, a rescaling is performed and those fields can be identified with geometric quantities of spacetime. Moreover, since at quantum level there is no relation between gauge space and spacetime, the spacetime is assumed to be Euclidean (a scenario that allows explicit computations for a QFT). This idea is indeed self consistent and can be applied to several gauge groups. In particular, the present work study the case of the affine group. However, due to the non-semisimplicity of the translational annex we are not able to write down an explicit action and thus we restrict ourselves to classical consequences of the model. On the other hand, we consider a general framework for the affine gauge theory that generates results that are independent of the specific dynamical equations. In fact, they depend exclusively on the topological features of a gauge theory and their description in terms of fibre bundles \cite{Kobayashi,Singer:1978dk,Daniel:1979ez,CottaRamusino:1985ad,Falqui:1985iu,Nakahara:1990th,Bertlmann:1996xk}.

By considering the framework developed in \cite{Sobreiro:2010ji} we are able to construct an affine gauge theory in a 4-dimensional spacetime in which the classical sector of the theory is a Riemann-Cartan geometrical gravity with extra matter fields. The path from a massless quantum gauge theory to a geometrical gravity is assumed to be mediated by mass parameters such as the Gribov parameter, always present in non-Abelian gauge theories with nontrivial topological sectors \cite{Singer:1978dk,Gribov:1977wm,Sobreiro:2005ec}. This main result is independent of the starting dynamics of the model, \emph{i.e.}, it is valid for any affine gauge theory.

This work is organized as follows: In Section 2 we provide the fundamental ideas concerning the proposed quantum gravity model. At the same Section we also construct the gauge theory for an affine group and show how it is related to the standard metric-affine gravities \cite{Hehl:1976my,Hehl:1994ue}. In Section 3 we show how the theory can in fact be contracted down to an orthogonal gauge theory with extra matter fields. Finally in Section 4 we display our conclusions.

\section{Quantum gravity as an Euclidean gauge theory}

A consistent gauge theory for gravity should encode both quantum and classical sectors. The classical regime is more or less well understood: It is a field theory whose dynamics determines the geometry of spacetime; the dynamical fields are the metric tensor and spacetime connection. To describe the quantum sector we propose that quantum gravity is a gauge theory in flat space, just like all other fundamental interactions. Thus, at high energies the quantum sector dominates and the model is a standard QFT with specific non-Abelian gauge symmetry embedded in flat space. Moreover, we introduce matter fields with dimension 1 which are translation algebra-valued\footnote{By algebra-valued we mean that the quantity takes values in the algebra of the group, in this case the group of translations, \emph{i.e.}, the quantity is expanded into the generators of the group. As a consequence its components will carry the corresponding group index.}\footnote{This requirement restricts the possible gauge groups to be considered.}. The existence of a mass scale ensures the possibility of mapping the matter fields into coframes and then identify the effective dynamical variables with geometry and thus to relate the model with gravity. Let us discuss the present proposition. To do so we enunciate the postulates of the model and their consequences.
\begin{itemize}
\item {\bf Flat spacetime}: \emph{The spacetime of the fundamental theory, {i.e.}, at very high energies, is a four dimensional Euclidean space} $\mathbb{R}^4$. This is the simplest choice to be made, so we made it. In fact, Euclidean spacetime is the proper scenario to make explicit computations in a QFT. Further, nonperturbative effects require Euclidean spacetime since Wick`s theorem is not known to be valid outside perturbation theory.

\item {\bf Gauge symmetry}: \emph{The quantum sector is described by a gauge theory with symmetry group denoted by} $\frak{G}$. At quantum level the gauge symmetry is totally independent of spacetime symmetries. Moreover, $\frak{G}$ is required to enjoy any morphism with spacetime in the sense that some topological invariants of the spacetime Euclidean structure should coincide with those of the gauge group. This last requirement is necessary to identify the gauge symmetry with the spacetime structure. Furthermore, we also require that $\frak{G}$ encodes as subgroup the group of translations $\mathbb{R}^4_g$ or at least that is related to it by forming a larger group. The index $g$ establishes that the space is the gauge space and is not related with spacetime.

\item {\bf Matter}: \emph{Massless matter fields algebra-valued in $\mathbb{R}^4_g$ carrying dimension 1 are introduced}. Those fields are required depending on the gauge group to be considered in order to avoid the intricacies that might emerge from groups that span non-flat gauge spaces.

\item {\bf Mass scales and geometry}: \emph{A mass scale $m$ should emerge dynamically from the quantum theory}. A mass scale is necessary in order to scale the UV and IR regimes. Once the energy decreases the mass scale dominates and throw the theory to a different behavior from the one at the UV scale. At this level, one can perform the rescaling of the matter fields and map them into a spacetime vierbein $e^a$ field and premetric $\frak{g}_{ab}$. Moreover, the gauge field should also be identified with the connection of a spacetime bundle (for instance the coframe bundle). Thus, the dynamics of the quantum theory are now dictating the spacetime structure at classical level. A metric tensor can then be defined
\begin{equation}
\frak{g}_{\mu\nu}=\frak{g}_{ab}e^a_\mu e^b_\nu\;.\label{g1}
\end{equation}
We remark that the most suitable candidate for the mass parameter is the so called Gribov parameter \cite{Gribov:1977wm,Sobreiro:2005ec,Dudal:2005na} associated with the improvement of the quantization of non-Abelian gauge theories.
\end{itemize}

The present program provides a self consistent way to define a quantum gravity theory in four dimensions. However, the formalization of the model is necessary in order to make explicit computations and predictions. The first step is to select the most suitable gauge group of the model. It turns out that the most general gauge group that can be fully mapped into a 4-dimensional spacetime is the 20 parameter affine group $A(4,\mathbb{R})$. The study of some properties of the respective gauge theory is the task of the rest of this work.

\subsection{Algebraic structure of the affine group}\label{ap1}

The affine group is defined as $A(d,\mathbb{R})\cong
GL(d,\mathbb{R})\ltimes\mathbb{R}^d$ where $GL(d,\mathbb{R})$ is the general linear group of $d\times d$
invertible real matrices and $\mathbb{R}^d$ stands for the group of
translations. Generically, we have for the Lie algebra decomposition
\begin{equation}
\left[gl,gl\right]\subseteq gl\;,\;\;\;\left[r,r\right]\subseteq\emptyset\;,\;\;\;\left[r,gl\right]\subseteq r\;,\label{alg000}
\end{equation}
where $gl$ is the algebra of the $GL(d,\mathbb{R})$ and $r$ the
algebra of the $\mathbb{R}^d$ translations. We remark that the affine group is not a semi-simple group due
to the translational sector. Further, the algebra decomposition
\eqref{alg000}, implies that the coset space $\mathbb{R}^d\cong
A(d,\mathbb{R})/GL(d,\mathbb{R})$ is a symmetric space and $GL(d,\mathbb{R})$ is a noncompact stability group of $A(d,\mathbb{R})$.

The general linear group can also be decomposed as $GL(d,\mathbb{R})\cong K(d)\otimes O(d)$ where $K(d)$ is the space defined from the symmetric part of the algebra of the general linear group. Formally, $K(d)$ is the coset space $K(d)\cong GL(d,\mathbb{R})/O(d)$ and, differently from $\mathbb{R}^d$, it does not form a group. The orthogonal group $O(d)$ is a maximal compact subgroup of the affine group as well as for the general linear group. The algebra decomposition can be described through
\begin{equation}
\left[o,o\right]\subseteq o\;,\;\;\;\left[k,k\right]\subseteq o\;,\;\;\;\left[k,o\right]\subseteq k\;,\label{coset0}
\end{equation}
where $o$ stands for the algebra of the orthogonal $O(d)$ group and
$k$ for the algebra of the coset space $K(d)$. From
(\ref{coset0}) one can infer that $K(d)$ forms a
symmetric space.

The elements of the affine group are here represented by exponentials according to $\mathcal{U}=\mathrm{e}^{\zeta(x)}=\mathrm{e}^{\alpha^a_{\phantom{a}b}(x)T^b_{\phantom{b}a}+\xi^a(x)P_a}$ where we are considering the algebraic notation of the adjoint representation of the affine group.  In fact, in this article, we are exclusively working in the adjoint representation. Thus, small Latin indices are vector indices of the $d$-dimensional differential manifold defined by the affine group, $a\in\{1,2,\ldots,d\}$. The parameter $\zeta$ is a $A(d,\mathbb{R})$ algebra-valued quantity $\zeta=\zeta^AS_A$ where $S_A$ represents the group generators in adjoint representation with $A\in\{1,2,\ldots,d(d+1)\}$. At the same level, $\alpha^a_{\phantom{a}b}$ and $\xi^a$ are algebra-valued parameters associated with the group element $U\in GL(d,\mathbb{R})$ and $u\in\mathbb{R}^d$, respectively, while $T^b_{\phantom{b}a}$ and $P_a$ the respective generators. Further, it is a general property of Lie groups that the element $\mathcal{U}$ can always be decomposed as $\mathcal{U}=uU$. Furthermore, under decomposition \eqref{coset0} the elements $U$ can be splitted according to $U=LC$, with $L=\mathrm{e}^{h^a_{\phantom{a}b}(x)\Sigma^b_{\phantom{b}a}}\in O(d)$ and $C=\mathrm{e}^{b^a_{\phantom{a}b}(x)\lambda^b_{\phantom{b}a}}\in K(d)$, where the notation speaks for itself.

\subsection{Affine gauge theory in Euclidean spacetime}

Following the construction of \cite{Sobreiro:2010ji,Daniel:1979ez,Nakahara:1990th,Bertlmann:1996xk,Trautman:1979cq,Trautman:1981fd} a gauge connection for the affine group, under the context of the previous section, arises from the principal bundle $A(4,\mathbb{R})\equiv\{A_r(4,\mathbb{R}),\mathbb{R}^4\}$ where the index $r$ stands for the rigid group and $\mathbb{R}^4$ the 4-dimensional Euclidean spacetime. The spacetime is then the base space while the affine group is the fiber as well as the structure group. A principal bundle of this form defines at each spacetime point a copy of the group, providing then a local character for the gauge group, \emph{i.e.}, the principal bundle is indeed the local affine group. In this bundle a gauge connection $Y=\omega+E$ emerges, where $\omega=\omega^a_{\phantom{a}b}T_a^{\phantom{a}b}$ is the general linear connection and $E=E^aP_a$ the translational one. A translation in the principal bundle $A(4,\mathbb{R})$, when restricted to a single fibre ($x\in\mathbb{R}^d$ is fixed), defines a gauge transformation
\begin{equation}
Y\longmapsto\mathcal{U}^{-1}\left(\mathrm{d}+Y\right)\mathcal{U}\;,\label{gauge02}
\end{equation}
where $\mathrm{d}$ is the space-time exterior derivative. At infinitesimal level the transformation \eqref{gauge02} reduces to $Y\longmapsto Y+\nabla\zeta$ where the covariant derivative is defined as $\nabla=\mathrm{d}+Y$ as long as we are considering the adjoint representation. Also at infinitesimal level the transformation \eqref{gauge02} decomposes as $\omega\longmapsto\omega+\mathcal{D}\alpha$ and $E\longmapsto E+\mathcal{D}\xi+E\alpha$, where the covariant derivative $\mathcal{D}$ is taken only with respect to the $GL(d,\mathbb{R})$ connection $\mathcal{D}=\mathrm{d}+\omega$. 

From the definition of the covariant derivative one can easily compute the field strength two-form $\nabla^2=\Theta=\Omega^a_{\phantom{a}b}T^b_{\phantom{b}a}+\Xi^aP_a$ where $\Omega^a_{\phantom{a}b}$ is the $GL(d,\mathbb{R})$ curvature
while $\Xi^a$ the translational one,
\begin{eqnarray}
\Omega^a_{\phantom{a}b}&=&\mathrm{d}\omega^a_{\phantom{a}b}-\omega_{\phantom{a}c}^a\omega^c_{\phantom{c}b}\;,\nonumber\\
\Xi^a&=&\mathrm{d}E^a-\omega_{\phantom{a}b}^aE^b\;=\;\nabla E^a\;=\;\mathcal{D}E^a\;.\label{curv02}
\end{eqnarray}
In despite of the fact that the field strength $\Theta$ is a covariant quantity under affine gauge transformations its components are not. In fact, at infinitesimal level it decomposes as $\Omega\longmapsto\Omega+\Omega\alpha$ and $\Xi\longmapsto\Xi+\Xi\alpha+\Omega\xi$.

The bundle $A(4,\mathbb{R})$ is very useful to define the correct gauge fields and their respective field strengths. However, as discussed in \cite{Sobreiro:2010ji}, it is a static bundle. To construct a dynamical theory one should migrate to an infinite dimensional principal bundle with the local gauge group $A(4,\mathbb{R})$ as fiber and the space of all independent gauge connections $\mathcal{Y}$ as base space \cite{Sobreiro:2010ji,CottaRamusino:1985ad,Falqui:1985iu,Nakahara:1990th,Bertlmann:1996xk}. By gauge independent connections we mean those that are not related to each other by a gauge transformation. Thus this principal bundle is denoted by $\mathbb{A}\equiv\{A(4,\mathbb{R}),\mathcal{Y}\}$. A typical fiber is the so called gauge orbit
\begin{equation}Y^{\mathcal{U}}=\mathcal{U}^{-1}\left(\mathrm{d}+Y\right)\mathcal{U}\;\Big|\;\;Y\in\mathcal{Y}\;.\label{orbit}
\end{equation}

A problem to be faced is the fact that the gauge space spanned by the local group $A(4,\mathbb{R})$ is not flat. Thus, extra degrees of freedom have to be considered in the theory. Those are represented by the matter fields $\sigma=\sigma^aP_a$ and $g=g^{ab}P_aP_b$. The fields $\sigma^a$ is a 1-form and possesses components with dimension 1. The field $g^{ab}$ also carries dimension 1 whereas it is a 0-form field double algebra-valued\footnote{By doble algebra-valued we mean that this quantity will carry two indices associated with the translation group. It is then expanded twice on the group generators.} in $\mathbb{R}^4_g$. Being matter fields both transform covariantly under gauge transformations. See \cite{Sobreiro:2010ji} for extra details.

The affine gauge theory here considered is then constructed with gauge fields $(\omega,E)$, matter fields $(\sigma,g)$ and no mass parameters. A consistent simple gauge invariant action should then be considered through the gauge covariant quantities $\Theta$, $T$ and $Q$ where the last two are the minimal couplings of the matter fields with the gauge fields
\begin{eqnarray}
T^a&=&\nabla\sigma^a=\mathcal{D}\sigma^a=\mathrm{d}\sigma^a-\omega^a_{\phantom{a}b}\sigma^b\;,\nonumber\\
Q^{ab}&=&\nabla g^{ab}=\mathcal{D}g^{ab}=\mathrm{d}g^{ab}-\omega^a_{\phantom{a}c}g^{cb}-\omega^b_{\phantom{b}c}g^{ca}\;.\label{mc1}
\end{eqnarray}
It is worth mention that, in the presence of mass parameters, a relation between $\Xi^a$ and $T^a$ can be enforced if an identification between $\sigma$ and $E$ is assumed. However, such association requires the introduction of an auxiliary field $\varphi^a$ in order to adjust the gauge field transformations between $E^a$ and $\sigma^a$, see \cite{Hehl:1994ue,Trautman:1979cq,Sobreiro:2010ji}. Anyhow, the affine group has a non-semisimple annex, the $\mathbb{R}^4_g$, that is characterized by the fact that there is no invariant Killing metric and thus disables a simple construction of a gauge invariant action \cite{Aldrovandi:1985gt}. Nevertheless, general linear algebra-valued quantities are allowed and can be used to define a theory on this sector (The most correct option is indeed to start from a $GL(4,\mathbb{R})$ gauge theory instead). We shall then skip the construction of an action and assume that an alternative method is successfully employed and study the consequences for the classical sector. Two remarks are in order: First that even in an affine gauge theory there are gauge invariant quantities and to construct them one should use a dual representation \cite{Sobreiro:2010ji} avoiding then the use of the metric tensor of the gauge space spanned by the affine group. In here, those degrees are represented by $\sigma$ and $g$ and are not directly related with the geometry of any space as long as their components are carrying  dimension 1. Second, the affine group has a compact subgroup and thus the respective gauge theory is plagued by Gribov ambiguities \cite{Singer:1978dk,Gribov:1977wm,Sobreiro:2005ec} implying that a mass parameter emerges in an improved quantization of the model, namely $m$. Notwithstanding, many other mass parameters might emerge as long as we consider non-Abelian gauge theories.

\subsection{Rescales and metric-affine gravity}

The idea behind a gravity gauge theory is that, after still unknown dynamical effects, the gauge space is identified space-time \cite{Mardones:1990qc,Zanelli:2005sa,Hehl:1994ue} and gravity as a geometric dynamical theory emerges. Considering, for instance, the Gribov parameter, one can rescale the fields $\sigma$ and $g$ together with the corresponding identification with vierbein and premetric of the cotangent spacetime bundle:
\begin{eqnarray}
\langle\sigma^a\rangle&=&me^a\;,\nonumber\\
\langle g^{ab}\rangle&=&m\frak{g}^{ab}\;,\label{id1}
\end{eqnarray}
and similar relations for the dual variables \cite{Sobreiro:2010ji}. As a consequence, expression \eqref{g1} is fulfilled and a geometric theory emerges, being associated with a gravity theory. The rescaling and identifications \eqref{id1} actually identify the gauge space with the cotangent spacetime bundle and thus say that the dynamics of the gauge theory dictates the geometry of the spacetime. In fact, the geometry that rises is the metric-affine geometry \cite{Hehl:1976my,Hehl:1994ue} with no symmetry breaking to the general linear group. To see this one can simply take a quick look at the field strengths and minimal couplings through the map \eqref{id1}. The field strength $\Omega$ is then mapped into the curvature while $\Xi$ is the translational curvature. Also, we have
\begin{eqnarray}
T\longmapsto m C\;,\nonumber\\
Q\longmapsto m N\;,
\end{eqnarray}
where $C$ is the Cartan Torsion 2-form and $N$ is the 1-form nonmetricity.

\section{Bundle contractions, Riemman-Cartan gravity and extra matter}

An alternative path is to consider the topological properties of the bundles $A(4,\mathbb{R})$ and $\mathbb{A}$. In fact, as discussed in detail in \cite{Sobreiro:2010ji}, the affine group has two contractible pieces $\mathbb{R}_g^4$ and $K(4)$ that can be reduced to define simpler and smaller principal bundles.

\subsection{Static bundle}

The contractible sectors $\mathbb{R}_g^4$ and $K(4)$ are both symmetric coset spaces, acording to \eqref{alg000} and \eqref{coset0}. As a consequence the $A(4,\mathbb{R})$ bundle can be contracted down to a smaller bundle \cite{Sobreiro:2010ji,Kobayashi,McInnes:1984kz}. Further, the connection on the reduced bundle is defined by the stability sector of the full connection \cite{Sobreiro:2010ji,Kobayashi,McInnes:1984kz}. In fact, the bundle contractions for the affine local group obey the following hierarchy
\begin{equation}
A(4,\mathbb{R})\rightarrow GL(4,\mathbb{R})\rightarrow O(4)\label{red1}
\end{equation}
\emph{i.e.}, the affine gauge theory naturally reduces to the orthogonal one. The major consequence comes from the static nature of this bundle, the connection $Y$ is also reduced to an orthogonal connection $w$ implying that all nonorthogonal degrees of freedom decouples from the bundle. As discussed in \cite{Sobreiro:2010ji,McInnes:1984kz} this result implies that every connection on $A(d,\mathbb{R})$ imposes a connection on $O(d)$. Now, if the bundle $A(4,\mathbb{R})$ is identified with the coframe bundle then the resulting spacetime geometry is the Rieman-Cartan one \cite{Sobreiro:2010ji,McInnes:1984kz}.

\subsection{Dynamical bundle}

The contraction \eqref{red1} induces a contraction of the dynamical bundle $\mathbb{A}$. However, in that case there are two main differences: The fact that we consider all possible gauge configurations implies that the full affine connection cannot be simply reduced to the orthogonal one. To see this one should look at the gauge orbit \eqref{orbit} which is also the fiber of the bundle. It means that the connection $Y$ is indeed part of the fundamental structure of the principal bundle and thus cannot simply decouple from it. The induced contraction on $\mathbb{A}$ can be performed by first contracting down the structure group $A(d,\mathbb{R})$ and then, by taking into account that all gauge connections and their equivalents were brought into the structure of the bundle $\mathbb{A}$ before the contraction and then analyze the respective effect on a typical fibre \eqref{orbit}. Second, as we shall see soon, the fact that the subgroup $\mathbb{R}^4_g$ is contractible will induce a matter character to $E$ that can be directly identified with the vierbein $e$. Thus, there is no need to introduce the matter field $\sigma$ at all. This second statement is important because simplifies the original setting of fields\footnote{We remark that this procedure is equivalent to assume the identification of $\mathbb{A}$ with a metric-affine gravity in which the contraction is performed afterwards. In this case $E$ and $e$ are identified by extra relations.} to $Y=(\omega,E)$ and $g$. This present approach slightly differs from the original developed in \cite{Sobreiro:2010ji}.

To perform the contraction we decompose the full affine connection according to $Y=w+q+E$ where $q$ is the $K(d)$ sector of the connection, \emph{i.e.}, $w=\omega^a_{\phantom{a}b}\Sigma^b_{\phantom{b}a}$ and $q=\omega^a_{\phantom{a}b}\lambda^b_{\phantom{b}a}$.  Let us take then a generic fibre \eqref{orbit} and contract the group space according to $\mathcal{U}\longrightarrow L$. Thanks to the symmetric nature of the coset spaces $\mathbb{R}^d$ and $K(d)$, the resulting decomposition is
\begin{eqnarray}
w^L&=&L^{-1}(\mathrm{d}+w)L\;,\nonumber\\
q^L&=&L^{-1}qL\;,\nonumber\\
E^L&=&L^{-1}E\;,\label{decomp03}
\end{eqnarray}
\emph{i.e.}, there are no mixing between the coset sectors and the orthogonal one. Moreover, for the matter sector, the contraction implies in $g^L=L^{-1}g\left(L^{-1}\right)^T$. Thus, the non-orthogonal sector of $Y$ abandon the geometric sector to migrate to the matter sector. Consequently, the metric-affine dynamical geometry described by $\mathbb{A}$ contracts down to a Riemann-Cartan one while the non-orthogonal components of the original connection are thrown into the matter sector. This effect is described by the contraction
\begin{equation}
\mathbb{A}\longrightarrow\mathbb{O}\oplus\mathcal{E}\oplus\mathcal{Q}\;,\label{contra03}
\end{equation}
where $\mathbb{O}\equiv\{O(4),\mathcal{L}\}$ is the dynamical principal bundle for an orthogonal theory. The space $\mathcal{L}\subset\mathcal{W}$ is the space of all independent orthogonal connections and $\mathcal{Q}=\mathcal{W}/\mathcal{L}$ is the space of the $K(4)$ connections. As an immediate consequence we can safely say that \emph{If one starts with a dynamical metric-affine gauge theory of gravity it can end up in a Riemann-Cartan theory with two extra matter fields} \cite{Sobreiro:2010ji}. In fact, the field strength decomposition yields
\begin{eqnarray}
\Omega^a_{\phantom{a}b}&=&R^a_{\phantom{a}b}+V^a_{\phantom{a}b}\;,\nonumber\\
\Xi^a&=&\mathrm{D}E^a-q^a_{\phantom{b}b}E^b\;,\label{decompcurv}
\end{eqnarray}
with $R^a_{\phantom{a}b}=\mathrm{d}w^a_{\phantom{a}b}-w^a_{\phantom{a}c}w^c_{\phantom{c}b}$ and $V^a_{\phantom{a}b}=\mathrm{D}q^a_{\phantom{a}b}-q^a_{\phantom{a}c}q^c_{\phantom{b}b}$ where $D=\mathcal{D}(w)$ is the orthogonal covariant derivative. Obviously, $R$ is the orthogonal field strength while $V$ is the orthogonal minimal coupling of $q$ with a nonlinear mater interaction term. The same interpretation follows for $E$ and its coupling $\Xi$. For the minimal coupling of $g$ it is achieved
\begin{equation}
Q^{ab}=\mathrm{D}g^{ab}-q^a_{\phantom{a}c}g^{cb}-q^b_{\phantom{b}c}g^{ca}\;,\label{mincoup01}
\end{equation}

\subsection{Rescales and Riemann-Cartan gravity}

To perform an identification with an induced geometry of spacetime we perform the mapping between $E$ and the vierbein together with $g$ and the premetric. To do so we have to keep in mind that the group space is now flat with a deformed rigid dynamical premetric. Thus
\begin{eqnarray}
\langle E^a\rangle&=&me^a\;,\nonumber\\
\langle g^{ab}\rangle&=&m\delta^{ab}+\gamma^{ab}\;,\label{id2}
\end{eqnarray}
where, again, $m$ is the Gribov parameter associated with the nontriviality of the orthogonal sector of the $\mathbb{A}$ and $\gamma^{ab}$ corresponds to the post-Riemannian sector of the original metric tensor of metric-affine gravity. Thus, $\delta^{ab}$ will work as the flat metric of cotangent space while $\gamma^{ab}$ will be a generic covariant matter field, symmetric in the group indices. In fact, from the transformation of $g$ under the affine group we obtain through contraction $\delta\longmapsto\delta$ and $\gamma\longmapsto L^{-1}\gamma\left(L^{-1}\right)^T$. The identification is then performed as
\begin{eqnarray}
\frak{g}_{\mu\nu}&=&\delta_{ab}e^a\otimes e^b\;,\nonumber\\
\frak{g}^{\mu\nu}&=&\delta^{ab}e_a\otimes e_b\;,\label{idx}
\end{eqnarray}
ensuring the Riemannian character of the spacetime. The effect of this identification into the $\mathbb{R}^4_g$ curvature is that $\Xi^a=m(C^a-q^a_{\phantom{b}b}e^b)$ while for the minimal coupling $Q$ it reduces to $Q^{ab}=\mathrm{D}\gamma^{ab}-2mq^{ab}-q^a_{\phantom{a}c}\gamma^{cb}-q^b_{\phantom{b}c}\gamma^{ca}$. Obviously, $\Xi$ is related to the Cartan torsion and a matter piece while $Q$ is a pure matter sector.

\section{Conclusions}\label{FINAL}

In this work we have proposed a suitable scenario for the construction of a quantum theory of gravity by stating the independence between the internal gauge space and the cotangent bundle of spacetime. In particular we considered the affine gauge group $A(4,\mathbb{R})$ and the starting spacetime as an Euclidean 4-dimensional one. This independence is ensured by considering the premetric as a genuine dimension 1 matter field interacting with the affine gauge connection. Although we have not provided a specific consistent quantum theory, we have studied some geometrical properties of the formalism. An important point is that the present framework allow for the compatibility of the principles of general relativity and those of quantum physics by assuming that both sectors shall not overlap, \emph{i.e.}, a consistent quantum model should be able to maintain the gauge internal space independent of the spacetime. Thus, dynamical effects associated to mass parameters, such as the Gribov parameter, are supposed to generate the identification between the cotangent space and gauge space. The theory is then automatically thrown in a geometric classical sector. We recall that this idea was already discussed in \cite{Sobreiro:2007pn} under the Palatini formalism and in \cite{Sobreiro:2010ji} already in the affine gauge model.

The main result of this article is the reduction of the metric-affine geometry to the Riemann-Cartan one. By considering the correct fibre bundle description of a gauge theory for the affine group, we have consistently contracted it to the orthogonal bundle. Thus, by decomposing the relevant fields into their orthogonal sector and the rest it is shown that we achieve a gauge theory for the orthogonal gauge group and extra matter fields. In fact, we have started with a theory with a gauge field $Y$ and a matter field $g$ and ended up in an orthogonal gauge theory with orthogonal spin-connection $w$, a vierbein field $e$ and matter fields $\gamma$ and $q$ associated with the original nonmetric degrees of a metric-affine gravity. We remark that, differently from \cite{Sobreiro:2010ji}, we have improved the formalization of the ideas of the quantum sector by reducing the number of starting fields as well as defining those fields as dimension 1 matter fields, which ensures the impossibility of geometric identification unless a mass parameter emerges.

In summary, in this work we provide a few ideas that goes on the direction of a possible quantum framework for a gauge theory of gravity as well as the formal analysis for the reduction of metric-affine geometry to the Riemann-Cartan one, independently of the starting field equations, resulting on the migration of the original nonmetric degrees to the matter set of fields.

\section*{Acknowledgements}

The authors express their gratitude to the Conselho Nacional de Desenvolvimento Cient\'{i}fico e Tecnol\'{o}gico\footnote{RFS is a level PQ-2 researcher under the program \emph{Produtividade em Pesquisa}, 304924/2009-1.} (CNPq-Brazil) for financial support. RFS is also partially supported, and gratefull for, by the Funda\c{c}\~ao Carlos Chagas Filho de Amparo \`a Pesquisa do Estado do Rio de Janeiro\footnote{Under the program \emph{Aux\'\i lio Instala\c{c}\~ao}, E-26/110.993/2009.} (FAPERJ) and the Pro-Reitoria de Pesquisa, P\'os-Gradua\c{c}\~ao e Inova\c{c}\~ao\footnote{Under the program \emph{Jovens Pesquisadores 2009}, project 305.} of the Universidade Federal Fluminense (Proppi-UFF).

\end{document}